\documentclass{PoS}
\usepackage{amsfonts,amssymb,amsmath,bm}
\usepackage{graphicx}

\newcommand{\be}{\begin{equation}}
\newcommand{\bea}{\begin{eqnarray}}
\newcommand{\ee}{\end{equation}}
\newcommand{\eea}{\end{eqnarray}}
\newcommand{\bpi}{\begin{picture}}
\newcommand{\bce}{\begin{center}}
\newcommand{\epi}{\end{picture}}
\newcommand{\ece}{\end{center}}

\def\NV{\bqq'}

\def\NP{ V}

\def\bqq{{\mathrm{I}\!\Gamma}}
\def\ff{B}

\def\bcj{J}

\title{Schwinger mechanism in QCD}

\ShortTitle{Schwinger mechanism in QCD}

\author{\speaker{Joannis Papavassiliou}\\
         Department of Theoretical Physics and IFIC,\\ 
        University of Valencia-CSIC,\\
        E-46100, Valencia, Spain.\\
        E-mail: \email{Joannis.Papavassiliou@uv.es}}

\abstract{The generation of a  
momentum-dependent gluon mass proceeds through a 
sophisticated implementation, at the level of the 
Schwinger-Dyson equation for the gluon propagator, 
of the Schwinger mechanism, whose 
central dynamical ingredient is the nonperturbative formation 
of longitudinally coupled massless bound-state excitations.
In addition to triggering the aforementioned mechanism, 
these excitations introduce poles in the various off-shell Green's functions 
of the theory, in such a way as to 
maintain the Slavnov-Taylor identities intact in the presence of 
massive gluon propagators, acting effectively  as composite 
Nambu-Goldstone bosons. 
In this work we focus on the dynamics leading to the 
actual formation of such bound states. Specifically, 
we derive and solve numerically an approximate version of the homogeneous 
Bethe-Salpeter equation governing the wave function  
of this special bound state. It is found  
that this integral equation admits physically meaningful non-trivial solutions, indicating that 
the QCD dynamics produce  one of  
the crucial ingredients required for the gauge-invariant generation 
of a gluon mass.}

\FullConference{International Workshop on QCD Green's Functions, Confinement and Phenomenology,\\
		September 05-09, 2011\\
		Trento Italy}

\begin{document}

\section{Introduction}

It is by now a well-established fact that 
large-volume lattice simulations in the Landau gauge yield a 
gluon propagator that reaches a finite non-vanishing value 
in the deep 
infrared~\cite{Cucchieri:2007md,Cucchieri:2009zt,Bogolubsky:2007ud,Bowman:2007du,Bogolubsky:2009dc,Oliveira:2009eh,arXiv:0912.0437}. 
Without a doubt, the most physical way of explaining this observed 
finiteness is to invoke the mechanism of dynamical gluon mass generation, first 
introduced in the seminal work of Cornwall~\cite{Cornwall:1981zr}, 
and subsequently studied in a series of articles~\cite{Aguilar:2006gr,Aguilar:2008xm,arXiv:1107.3968}. 
In this picture  
the fundamental Lagrangian of the Yang-Mills theory (or that of QCD) remains unaltered, and   
the generation of the gluon mass takes place dynamically, 
through the well-known Schwinger 
mechanism~\cite{Schwinger:1962tn,Schwinger:1962tp,Jackiw:1973tr,Jackiw:1973ha,Cornwall:1973ts,Eichten:1974et,Poggio:1974qs},  
without violating any of the underlying symmetries
(for related contributions and alternative approaches, see, e.g., 
\cite{Fischer:2008uz,Kondo:2009gc,RodriguezQuintero:2010ss,arXiv:1007.2583,Szczepaniak:2010fe,Oliveira:2010xc,arXiv:1106.3240,Pennington:2011xs}).

The way how 
the Schwinger mechanism generates a  mass for the gauge boson (gluon)  
can be seen most directly at the level of its 
inverse propagator, 
$\Delta^{-1}({q^2})=q^2 [1 + i {\bm \Pi}(q^2)]$, where ${\bm \Pi}(q)$ 
is the dimensionless vacuum polarization.
According to Schwinger's fundamental observation, 
if ${\bm \Pi}(q^2)$ 
develops a pole at zero momentum transfer ($q^2=0$), then the 
vector meson acquires a mass, even if the gauge symmetry 
forbids a mass term at the level of the fundamental Lagrangian.
Indeed, if ${\bm \Pi}(q^2) = m^2/q^2$, then (in Euclidean space)
\mbox{$\Delta^{-1}(q^2) = q^2 + m^2$}, and so 
the vector meson becomes massive, $\Delta^{-1}(0) = m^2$, 
even though it is massless in the absence of interactions 
($g=0$, ${\bm \Pi} =0$)~\cite{Jackiw:1973tr,Jackiw:1973ha}.

The key assumption when invoking the Schwinger mechanism in Yang-Mills theories, 
such as QCD, is that the required  poles   may be produced 
due to purely dynamical reasons; specifically, one assumes that, for sufficiently 
strong binding, 
the mass of the appropriate bound state may be reduced to zero~\cite{Jackiw:1973tr,Jackiw:1973ha,Cornwall:1973ts,Eichten:1974et,Poggio:1974qs}. 
In addition to triggering the Schwinger mechanism, 
these massless composite excitations are crucial for preserving gauge invariance. 
Specifically, the presence of massless poles  in the off-shell interaction vertices 
guarantees that the Ward identities (WIs) and Slavnov Taylor identities (STIs) of the theory 
maintain exactly the same form before and after mass generation (i.e. when the 
the massless propagators appearing in them are replaced by massive ones)
~\cite{Cornwall:1981zr,Eichten:1974et,Poggio:1974qs,arXiv:1107.3968}.
Thus, these excitations act like dynamical Nambu-Goldstone scalars,
displaying, in fact, all their typical characteristics, such as 
masslessness, compositeness, and longitudinal coupling; 
note, however, that they differ from Nambu-Goldstone bosons 
as far as their origin is concerned, since they are not associated 
with the spontaneous breaking of any global symmetry~\cite{Cornwall:1981zr}.
Finally, every such Goldstone-like scalar, ``absorbed'' by
a gluon in order to acquire a mass,  
is expected to actually cancel out of the $S$-matrix   
against other massless poles or due to current conservation~\cite{Jackiw:1973tr,Jackiw:1973ha,Cornwall:1973ts,Eichten:1974et,Poggio:1974qs}.

The main purpose of this presentation  is to report on recent work~\cite{arXiv:1110.2633}, 
where the central assumption of  
the dynamical scenario outlined above, namely the possibility of actual formation 
of such massless excitations, has been examined. 
Specifically,  
the entire mechanism of gluon mass generation hinges on the 
appearance of massless poles inside the nonperturbative three-gluon vertex, 
which enters in the Schwinger Dyson equation (SDE) governing the gluon propagator. 
These poles correspond to the propagator of the scalar massless excitation, 
and interact with a pair of gluons through a very characteristic proper vertex, 
which, of course, must be non vanishing, or else the entire construction is invalidated.
The way to establish the existence of this latter vertex  is by finding non-trivial solutions
to the homogeneous Bethe-Salpeter equation (BSE) that it satisfies. 

\section{Basic concepts}

The full gluon propagator 
$\Delta^{ab}_{\mu\nu}(q)=\delta^{ab}\Delta_{\mu\nu}(q)$ in the Landau gauge is defined as
\be
\Delta_{\mu\nu}(q)=- i P_{\mu\nu}(q)\Delta(q^2) \,,
\label{prop}
\ee 
where
\be
P_{\mu\nu}(q)=g_{\mu\nu}- \frac{q_\mu q_\nu}{q^2} \,,
\ee
is the usual transverse projector, 
and the scalar cofactor $\Delta(q^2)$  
is related to the (all-order) gluon self-energy $\Pi_{\mu\nu}(q)=P_{\mu\nu}(q)\Pi(q^2)$  through
\be
\Delta^{-1}({q^2})=q^2+i\Pi(q^2).
\label{defPi}
\ee
One may define the dimensionless vacuum polarization ${\bm \Pi}(q^2)$ 
by setting  $\Pi(q^2) = q^2 {\bm \Pi}(q^2)$ so that (\ref{defPi}) becomes 
\be
\Delta^{-1}({q^2})=q^2 [1 + i {\bm \Pi}(q^2)] \,.
\label{defvp}
\ee

Alternatively, one may define the gluon dressing function $J(q^2)$ as 
\be
\Delta^{-1}({q^2})=q^2 J(q^2) \,.
\label{defJ}
\ee
In the presence of a dynamically generated mass, the natural form of $\Delta^{-1}(q^2)$ is given by 
(Euclidean space) 
\be
\Delta^{-1}(q^2) =q^2 J(q^2) + m^2(q^2) \,,
\label{defm}
\ee
where the first term corresponds to the ``kinetic term'', or ``wave function'' contribution, 
whereas the second is the (positive-definite) momentum-dependent mass.
If one insists on maintaining the form of (\ref{defJ}) by explicitly factoring out a $q^2$, then   
\be
\Delta^{-1}({q^2})=q^2  \left[J(q^2) + \frac{m^2(q^2)}{q^2}\right]\,,
\label{defJm}
\ee
and the presence of the pole, with residue given by $m^2(0)$, becomes manifest. 
\begin{figure}[!t]
\begin{center}
\includegraphics[scale=0.8]{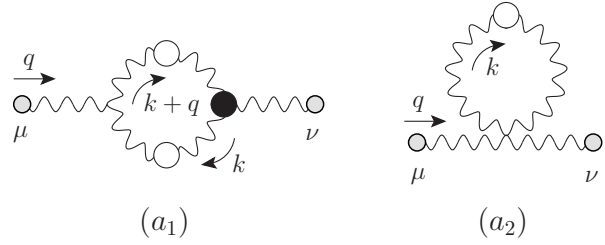}
\caption{\label{gSDE} The ``one-loop dressed'' gluon contribution to the  PT-BFM gluon self-energy. 
White (black) circles denote fully dressed propagators (vertices); 
a gray circle attached to the external legs indicates that they are background gluons. 
Within the PT-BFM framework these 
two diagrams constitute a transverse subset of the full gluon SDE.}
\end{center}
\end{figure}

The Schwinger mechanism
is integrated into the SDE of the gluon propagator through the form of the three-gluon vertex.
In particular,  
a crucial condition for the realization of the gluon mass generation scenario  
is the existence of a special vertex, to be denoted by $V_{\alpha\mu\nu}(q,r,p)$,  
which must be completely {\it longitudinally coupled}, 
i.e. must satisfy 
\be
P^{\alpha'\alpha}(q) P^{\mu'\mu}(r) P^{\nu'\nu}(p) V_{\alpha\mu\nu}(q,r,p)  = 0 \,.
\label{totlon}
\ee 

The role of the vertex $V_{\alpha\mu\nu}(q,r,p)$ is instrumental for maintaining gauge invariance, 
given that 
the massless poles that it must contain in order to trigger the Schwinger mechanism, 
act, at the same time, as composite, longitudinally coupled Nambu-Goldstone bosons. 
Specifically, in order to preserve the gauge invariance of the theory in the presence of masses, 
the vertex $V_{\alpha\mu\nu}(q,r,p)$ must be added to the 
conventional (fully-dressed) three-gluon vertex $\bqq_{\alpha\mu\nu}(q,r,p)$, giving rise 
to the new full vertex,  $\NV_{\alpha\mu\nu}(q,r,p)$, defined as  
\be
\NV_{\alpha\mu\nu}(q,r,p) = \bqq_{\alpha\mu\nu}(q,r,p) +V_{\alpha\mu\nu}(q,r,p)\,.
\label{NV}
\ee
Gauge invariance remains intact because $\NV$
satisfies the same WI (or STI) as $\bqq$ before, but now replacing the gluon propagators appearing on their
rhs by massive ones; schematically, $\Delta^{-1} \to \Delta_m^{-1}$, where the former denotes  
the propagator given in (\ref{defJ}), while the latter that of  (\ref{defm}).

To see this in detail, let us employ the formalism 
provided by the  synthesis of the pinch technique (PT)~\cite{Cornwall:1981zr,Cornwall:1989gv,Binosi:2009qm}
with the background field method (BFM)~\cite{Abbott:1980hw}. 
In this framework, the natural quantity to consider is the vertex $BQQ$, to be denoted by $\bqq_{\alpha\mu\nu}(q,r,p)$,
connecting a background gluon ($B$) with two quantum gluons ($Q$).
With the Schwinger mechanism turned off, this vertex satisfies the WI
\be
q^\alpha\bqq_{\alpha\mu\nu}(q,r,p)=p^2\bcj(p^2)P_{\mu\nu}(p)-r^2\bcj(r^2)P_{\mu\nu}(r)\,,
\label{STI}
\ee
when contracted with respect to the momentum of the background gluon.
Then,  gauge invariance requires that 
\be
q^\alpha V_{\alpha\mu\nu}(q,r,p)= m^2(r^2)P_{\mu\nu}(r) - m^2(p^2)P_{\mu\nu}(p) \,,
\label{winp}
\ee
so that, after turning the Schwinger mechanism on,  the corresponding WI satisfied by $\NV$ would read  
\bea
q^{\alpha}\NV_{\alpha\mu\nu}(q,r,p) &=& 
q^{\alpha}\left[\bqq(q,r,p) + \NP(q,r,p)\right]_{\alpha\mu\nu}
\nonumber\\
&=& [p^2\bcj (p^2) -m^2(p^2)]P_{\mu\nu}(p) - [r^2\bcj (r^2) -m^2(r^2)]P_{\mu\nu}(r)
\nonumber\\
&=& \Delta^{-1}_m({p^2})P_{\mu\nu}(p) - \Delta^{-1}_m({r^2})P_{\mu\nu}(r) \,,
\label{winpfull}
\eea
which is indeed the identity in Eq.~(\ref{STI}), with the aforementioned replacement 
\mbox{$\Delta^{-1} \to \Delta_m^{-1}$} enforced. 
The remaining STIs, triggered when contracting $\NV_{\alpha\mu\nu}(q,r,p)$ with respect to the other two legs  
are realized in exactly the same fashion.

The next step is to insert $\NV_{\alpha\mu\nu}(q,r,p)$ into the SDE equation satisfied by the 
gluon propagator, see Fig.~\ref{gSDE}.  
Then, a rather elaborate analysis~\cite{arXiv:1107.3968}  gives rise to an integral equation for 
the momentum-dependent gluon mass, 
of the type
\be
m^2(q^2) = \int_{k}\, m^2(k^2)\, K(q,k)\,,
\ee
where the kernel $K$ survives the $q \to 0$ limit, i.e., 
$\lim_{q \to 0}\, K(q,k) \neq 0$,  
precisely because it includes the term $1/q^2$ contained inside $V_{\alpha\mu\nu}(q,r,p)$.

\section{Structure of the pole vertex}

\begin{figure}[ht]
\center{\includegraphics[scale=0.5]{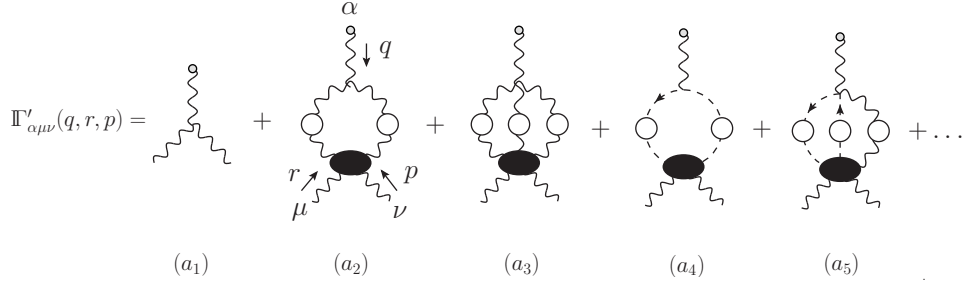}}
\caption{The SDE for the $BQQ$ vertex which connects a background gluon ($B$) with two quantum gluons ($Q$).}
\label{FullvertexSDE}
\end{figure}

The main characteristic of the vertex $V$, 
which sharply differentiates it from ordinary vertex contributions, 
is that it contains massless poles, originating from the 
contributions of bound-state excitations. 
Specifically, all terms of the vertex $V$ are proportional to $1/q^2$, $1/r^2$,  $1/p^2$, 
and products thereof. 
Such dynamically generated poles are to be 
clearly distinguished from poles related to  
ordinary massless propagators, associated with elementary fields in the original Lagrangian.

To see how such poles enter into the vertex, 
let us focus on the general structure of the SDE for the $BQQ$ vertex.  
With the Schwinger mechanism turned off, the various 
multiparticle kernels appearing in this SDE 
have a complicated skeleton expansion 
(not shown here), 
but their common characteristic is that they are one-particle-irreducible with respect to 
cuts in the direction of the momentum $q$ 

\begin{figure}[!t]
\center{\includegraphics[scale=0.5]{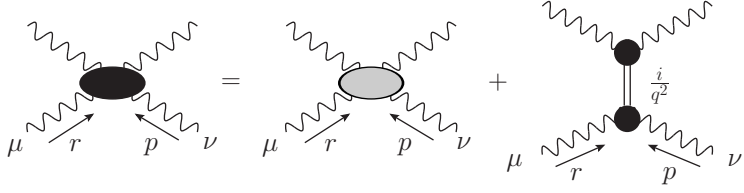}}
\caption{The gray kernel (regular part with respect to $q$, and the 
composite massless excitation in the $q$-channel.}
\label{Fullkernel}
\end{figure}

\begin{figure}[!t]
\center{\includegraphics[scale=0.45]{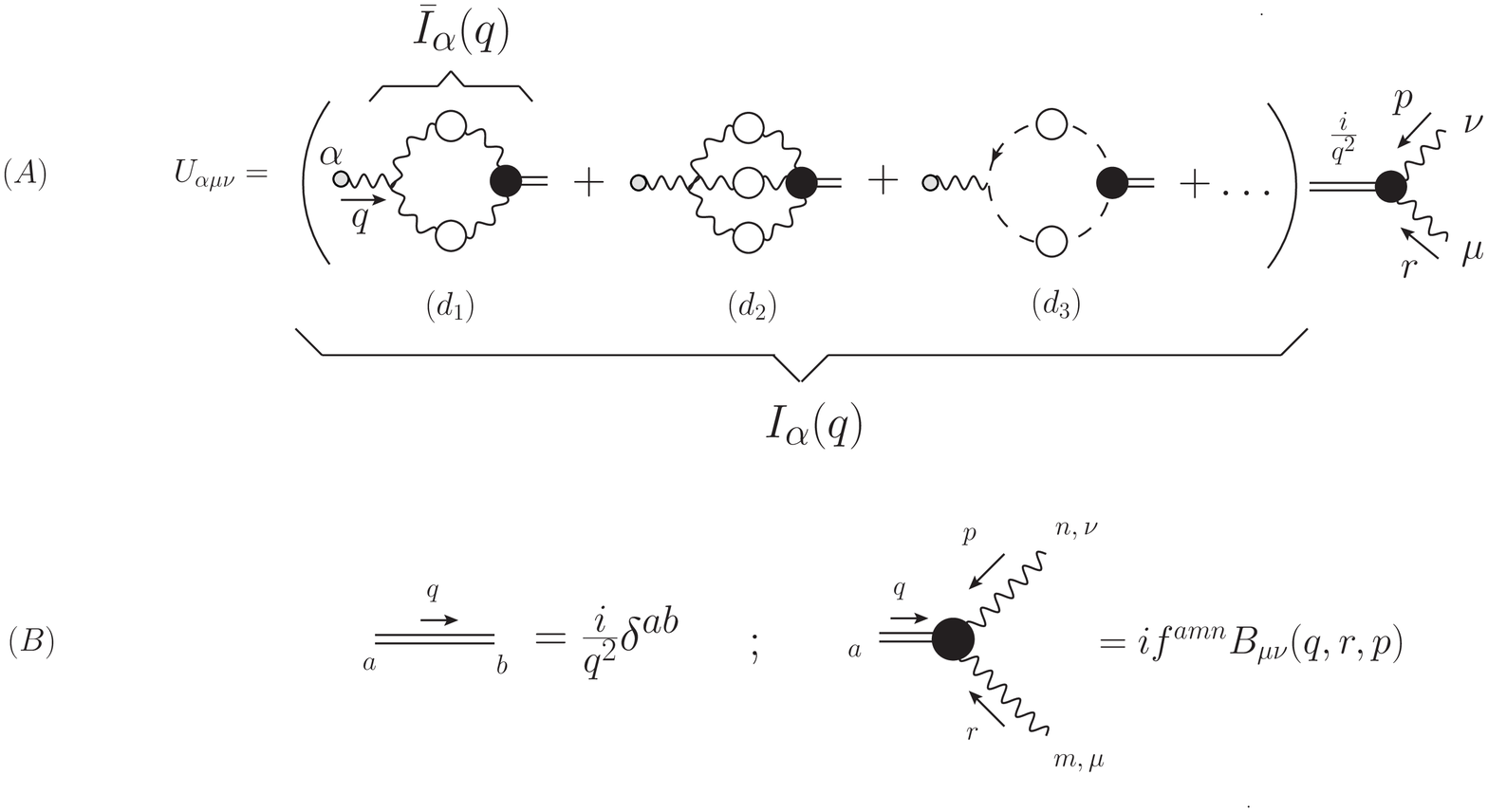}}
\caption{{\bf(A)} The vertex $U_{\alpha\mu\nu}$ is composed of three main ingredients: 
the transition amplitude, $I_{\alpha}$, which 
mixes the gluon with a massless excitation, 
the propagator of the  massless excitation, and the (massless excitation)--(gluon)--(gluon) vertex. 
{(\bf B)} The Feynman rules (with color factors included) for (i) the propagator of the massless excitation 
and (ii) the ``proper vertex function'', or, ``bound-state wave function'', $B_{\mu\nu}$.}  
\label{Uexpansion}
\end{figure}

When the Schwinger mechanism is turned on, 
the structure of the kernels is modified by the presence of composite  
massless excitation, described by a propagator of the type $i/q^2$, as shown in Fig.~\ref{Fullkernel}.
The sum of such dynamical terms, coming from all multiparticle kernels, 
shown in Fig.~\ref{Uexpansion}, constitutes a characteristic part of the vertex $V$, 
to be denoted by $U$ in  Eq.~(\ref{VU}),   
namely the part that contains at least a massless propagator $i/q^2$. 
The remaining parts, to be denoted by $R$,
contain massless excitations in the other two channels, 
namely $r_{\mu}/r^2$ and $p_{\nu}/p^2$ (but no $q_\alpha/q^2$), and are not relevant for the purposes 
of this presentation. Thus,
\be
V_{\alpha\mu\nu}(q,r,p) =  U_{\alpha\mu\nu}(q,r,p) + R_{\alpha\mu\nu}(q,r,p)  \,,
\label{VRU}
\ee
with 
\be
U_{\alpha\mu\nu}(q,r,p)= q_\alpha \bigg(V_1g_{\mu\nu}+V_2q_{\mu}q_{\nu}+V_3p_{\mu}p_{\nu}+
V_4r_{\mu}q_{\nu}+V_5r_{\mu}p_{\nu}\bigg)\,,
\label{VU}
\ee
where the $V_i$ are form factors depending on the various momenta.

At this point we can make the nonperturbative pole  
manifest, and cast ${U}_{\alpha\mu\nu}(q,r,p)$ in the form of Fig.~\ref{Uexpansion}, 
by setting
\be
{U}_{\alpha\mu\nu}(q,r,p) = I_{\alpha}(q)\left(\frac{i}{q^2} \right) \ff_{\mu\nu}(q,r,p) \,, 
\label{VwB}
\ee
where the nonperturbative quantity 
\be
\ff_{\mu\nu}(q,r,p) = B_1g_{\mu\nu}+B_2q_{\mu}q_{\nu}+B_3p_{\mu}p_{\nu}+B_4r_{\mu}q_{\nu}+B_5r_{\mu}p_{\nu} \,,
\ee
is the effective vertex 
describing the interaction between 
the massless excitation and two gluons.  
$\ff_{\mu\nu}(q,r,p)$ is to be identified with the ``bound-state wave function'' (or 
``BS wave function'')  
of the two-gluon bound-state shown in Fig.~\ref{Fullkernel}, which,  
as we will see shortly, satisfies a homogeneous BSE. 
In addition, ${i}/{q^2}$ is the 
propagator of the scalar massless excitation. Finally,   
$I_{\alpha}(q)$ is the (nonperturbative) transition 
amplitude introduced in Fig.~\ref{Uexpansion}, allowing the 
mixing between a gluon and the massless excitation; note that  
the imaginary factor ``$i$'' from the Feynman rule in Fig.~\ref{Fullkernel} 
is absorbed into the definition of $I_{\alpha}(q)$.


Evidently, by Lorentz invariance,  
\be
I_{\alpha}(q) = q_{\alpha} I(q)\,,
\label{qI}
\ee 
and the scalar cofactor, to be referred to as the ``transition function'', is simply given by 
\be
I(q) = \frac{q^{\alpha}{I}_{\alpha}(q)}{q^2} \,,
\label{trf}
\ee 
so that
\be
{V}_j (q,r,p)  = I(q) \left(\frac{i}{q^2} \right) \ff_j(q,r,p) \, ; \quad j=1,\dots,5 \,. 
\label{pole}
\ee

Note that, due to Bose symmetry with respect to the interchange 
$\mu \leftrightarrow \nu$ and  $p \leftrightarrow r$, we must have
\be
\ff_{1,2}(q,r,p) = - \ff_{1,2}(q,p,r)\,,
\label{anti}
\ee
which implies that 
\be
\ff_{1,2}(0,-p,p) = 0 \,.
\label{ant0}
\ee

\section{Gluon mass and the BS wave-function: an exact relation}

The WI of Eq~(\ref{winp}) furnishes an  
exact relation between the dynamical gluon mass, the transition amplitude at zero momentum transfer, 
and the form factor $B_1$. 
Specifically, contracting both sides of the WI with 
two transverse projectors, one obtains,
\be
P^{\mu'\mu}(r) P^{\nu'\nu}(p) q^\alpha V_{\alpha\mu\nu}(q,r,p) = [m^2(r)-m^2(p)] P^{\mu'}_\sigma(r) P^{\sigma\nu'}(p) \,.
\ee
On the other hand, contracting the full expansion of the vertex (\ref{VwB}) by these transverse projectors
and then contracting the result with the momentum of the background leg, we get 
\be
q^\alpha P^{\mu'\mu}(r) P^{\nu'\nu}(p) V_{\alpha\mu\nu}(q,r,p) = iI(q)[B_1g_{\mu\nu}+B_2q_\mu q_\nu] P^{\mu'\mu}(r) P^{\nu'\nu}(p) \,,
\ee
where the relation of Eq~(\ref{pole}) has been used. Thus, equating both results, one arrives at 
\be
i I(q) B_1(q,r,p) = m^2(r)-m^2(p) \,, \,\,\,\,\,\,\,\,\,\, B_2(q,r,p) = 0 . 
\label{factor1}
\ee
The above relations, together  with those of Eq.~(\ref{pole}),
determine exactly the form factors $V_1$ and $V_2$ of the vertex $V_{\alpha\mu\nu}$, namely
\be
V_1(q,r,p) = \frac{m^2(r)-m^2(p)}{q^2} \,, \,\,\,\,\,\,\,\,\,\,  V_2(q,r,p) = 0 . 
\ee

We will now carry out the Taylor expansion of both sides of Eq~(\ref{factor1}) in the limit \mbox{$q\rightarrow0$}.
To that end, let consider the Taylor expansion of a 
function \mbox{$f(q,r,p)$} around \mbox{$q=0$} (and \mbox{$r=-p$}). In general we have 
\begin{equation}
f(q,-p-q,p)=f(-p,p)+ [2(q\cdot p) + q^2 ]f'(-p,p)+2(q\cdot p)^2 f''(-p,p)+{\cal O}(q^3)\,,
\label{Taylor}
\end{equation}
where the prime denotes differentiation with respect to $(p+q)^2$ and subsequently taking the 
limit $q\to 0$, i.e. 
\begin{equation}
f'(-p,p) \equiv \, \lim_{q\to 0} \left\{ \frac{\partial f(q,-p-q,p)}{\partial\, (p+q)^2} \right\} \,.
\label{Der}
\end{equation}
Now, if the function is antisymmetric under $p \leftrightarrow r$, as happens with 
the form factors $\ff_{1,2}$, then $f(-p,p) = 0$; thus, for the case of the form factors 
in question, the Taylor expansion is ($i=1,2$)
\begin{equation}
\ff_i(q,-p-q,p)= [2(q\cdot p) + q^2] \ff'_i(-p,p) + 2(q\cdot p)^2 \ff''_i(-p,p)  +  {\cal O}(q^3) \,.
\label{TaylorB}
\end{equation}

Using Eq~(\ref{TaylorB}), and the corresponding expansion for the rhs, 
\be
m^2(r)-m^2(p) = m^2(q+p)-m^2(p) = 2(q\cdot p)[m^2(p)]' + {\cal O}(q^2) \quad,
\ee
assuming that the $I(0)$ is finite,
and equating the coefficients in front of $(q\cdot p)$, we arrive at  (Minkowski space)
\be
[m^2(p)]' =  i I(0) B'_1(p) \,. 
\label{massrelation}
\ee
Note that this is an exact relation,
whose derivation relies  only on the 
WI and  Bose-symmetry that $V_{\alpha\mu\nu}(q,r,p )$ 
satisfies, as captured by Eq.~(\ref{winp}) and Eq.~(\ref{ant0}), respectively.

\section{\label{bse}The Bethe-Salpeter equation}

\begin{figure}[!t]
\center{\includegraphics[scale=0.6]{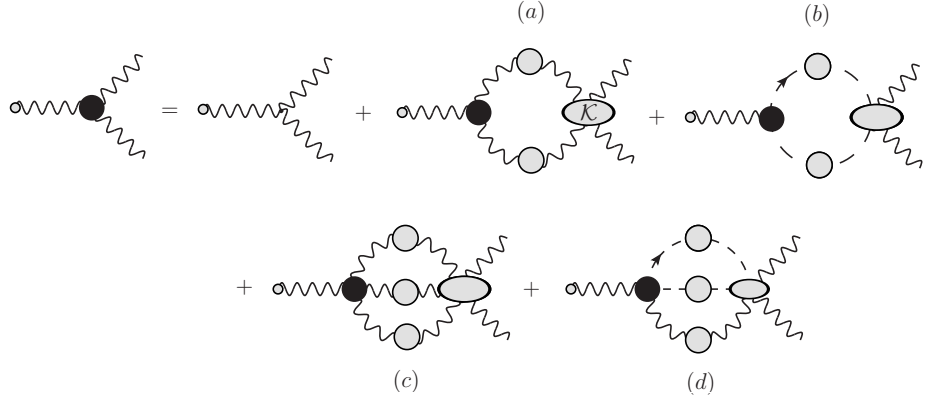}}
\caption{The complete BSE for the full three gluon vertex $\NV_{\alpha\mu\nu}(q,r,p)$.}
\label{Fig6}
\end{figure}

As has become clear in the previous section, 
the existence of $\ff'_1$ is of paramount 
importance for the mass generation mechanism envisaged here; essentially, the question 
boils down to whether or not   
the dynamical formation of a massless 
bound-state excitation of the type postulated above is possible.
As is well-known, in order to establish the 
existence of such a bound state one must 
{\bf (i)} derive the appropriate BSE for the 
corresponding bound-state wave function, $B_{\mu\nu}$,  
(or, in this case, its derivative),   
and {\bf (ii)} find non-trivial solutions for this integral equation.

The starting point is the BSE 
for the vertex $\NV_{\alpha\mu\nu}(q,r,p)$, shown in Fig.~\ref{Fig6}.
Note that, unlike the corresponding SDE of  Fig.~\ref{FullvertexSDE}, the vertices where the 
background gluon is entering (carrying momentum $q$) are now fully dressed.
 As a consequence, the corresponding multiparticle kernels appearing in   Fig.~\ref{Fig6} 
are different from those 
of the SDE. 

\begin{figure}[!t]
\center{\includegraphics[scale=0.5]{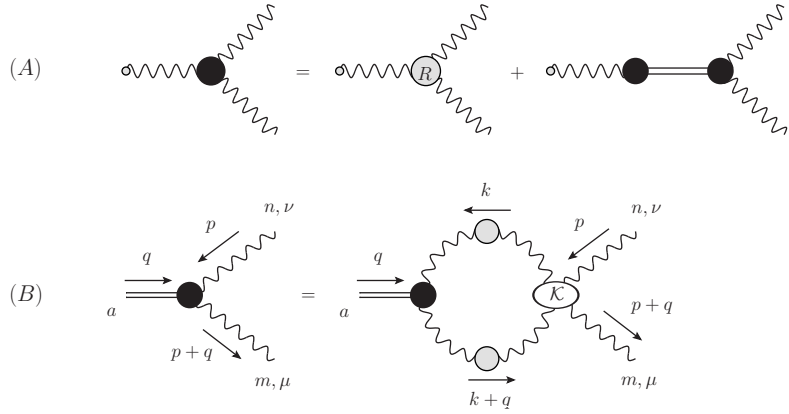}}
\caption{{\bf(A)} The separation of the vertex in regular and pole parts.
{\bf(B)} The BSE for the bound-state wave function $B_{\mu\nu}$.}
\label{RegPole}
\end{figure}

The general methodology of how to isolate from the BSE shown in Fig.~\ref{Fig6} 
the corresponding dynamical equation for 
the quantity $B_{\mu\nu}$ has been explained in~\cite{Jackiw:1973ha,Poggio:1974qs}. 
Specifically, one separates on both sides of the BSE equation 
each vertex (black circle) into two parts,  
a ``regular'' part and another containing a pole $1/q^2$; this separation 
is shown schematically in Fig.~\ref{RegPole}.
Then, omitting all other vertices, and the possible poles they too may have, 
the BSE for $B_{\mu\nu}(q,r,p)$
is obtained simply by equating the pole parts on both sides; 
specifically, [see Fig.~\ref{Fig6}]
\begin{equation}
B_{\mu\nu}^{amn} = 
\int_k B_{\alpha\beta}^{abc}\Delta^{\alpha\rho}_{br}(k+q)\Delta^{\beta\sigma}_{cs}(k){\cal K}_{\sigma\nu\mu\rho}^{snmr} \,.
\label{BS}
\end{equation}

\begin{figure}[!t]
\center{\includegraphics[scale=0.5]{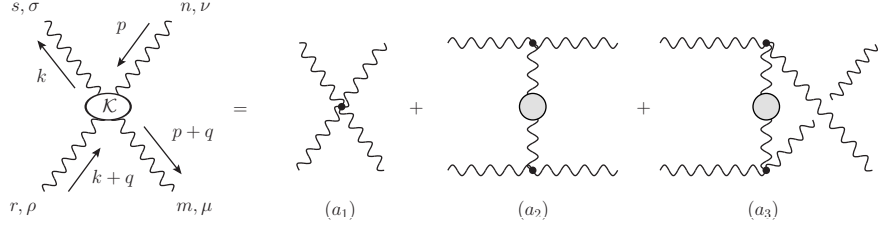}}
\caption{The Feynman diagrams considered for the BS kernel. 
The interaction vertices are approximated by their tree level values, 
while the internal gluon propagators are fully dressed.}
\label{Fig9}
\end{figure}

We will next approximate the  four-gluon BS kernel ${\cal K}$ by the 
lowest-order set of diagrams shown in Fig.~\ref{Fig9}, where the vertices are bare, while 
the internal gluon propagators are fully dressed.
Going to Euclidean space, we define $x\equiv p^2$, $y\equiv k^2$, and $z\equiv (p+k)^2$; 
then, after appropriate Taylor expansion, and use of the fact that $B_2=0$ [see Eq.~(\ref{factor1})],
the BSE becomes
\begin{eqnarray}\label{34}
B'_1(x) &=& -\frac{\alpha_s C_A}{12\pi^2}\int_0^\infty dy yB'_1(y)\Delta^2(y)\nonumber \\
&& \sqrt{\frac{y}{x}}\int_0^\pi d\theta\sin^4\theta\cos\theta\bigg[z+10(x+y)+\frac{1}{z}(x^2+y^2+10xy)\bigg]\Delta(z)\,.
\label{euclideanBS}
\end{eqnarray}

As a further simplification, we approximate the gluon propagator $\Delta(z)$ 
appearing in the BSE of (\ref{euclideanBS}) [but not the $\Delta^2(y)$]
by its tree level value, that is, $\Delta(z)=1/z$. Then, the angular integration may be carried out 
exactly, yielding  

\be
B'_1(x) = \frac{\alpha_s C_A}{24\pi}\bigg\lbrace\int_0^x dyB'_1(y)\Delta^2(y)\frac{y^2}{x}
\bigg(3+\frac{25}{4}\frac{y}{x}-\frac{3}{4}\frac{y^2}{x^2}\bigg)
+\int_x^\infty dyB'_1(y)\Delta^2(y)y\bigg(3+\frac{25}{4}\frac{x}{y}-\frac{3}{4}\frac{x^2}{y^2}\bigg)\bigg\rbrace \,.
\label{weakBS}
\ee

\section{Numerical analysis}
Next we discuss the numerical solutions for 
Eq.~(\ref{weakBS}) for arbitrary values of $x$. 
Evidently, the main ingredient entering into its kernel is
the nonperturbative gluon propagator, $\Delta(q)$. In order to explore the sensitivity of 
the solutions on the details of $\Delta(q)$, 
we  will employ three infrared-finite forms, to be denoted by $\Delta_1(q)$, $\Delta_2(q)$, 
and $\Delta_3(q)$, focusing on their differences  in the intermediate and asymptotic 
regions of momenta.

\begin{figure}[!t]
\begin{minipage}[b]{0.55\linewidth}
\noindent
\centering
\hspace{-1cm}
\includegraphics[scale=0.55]{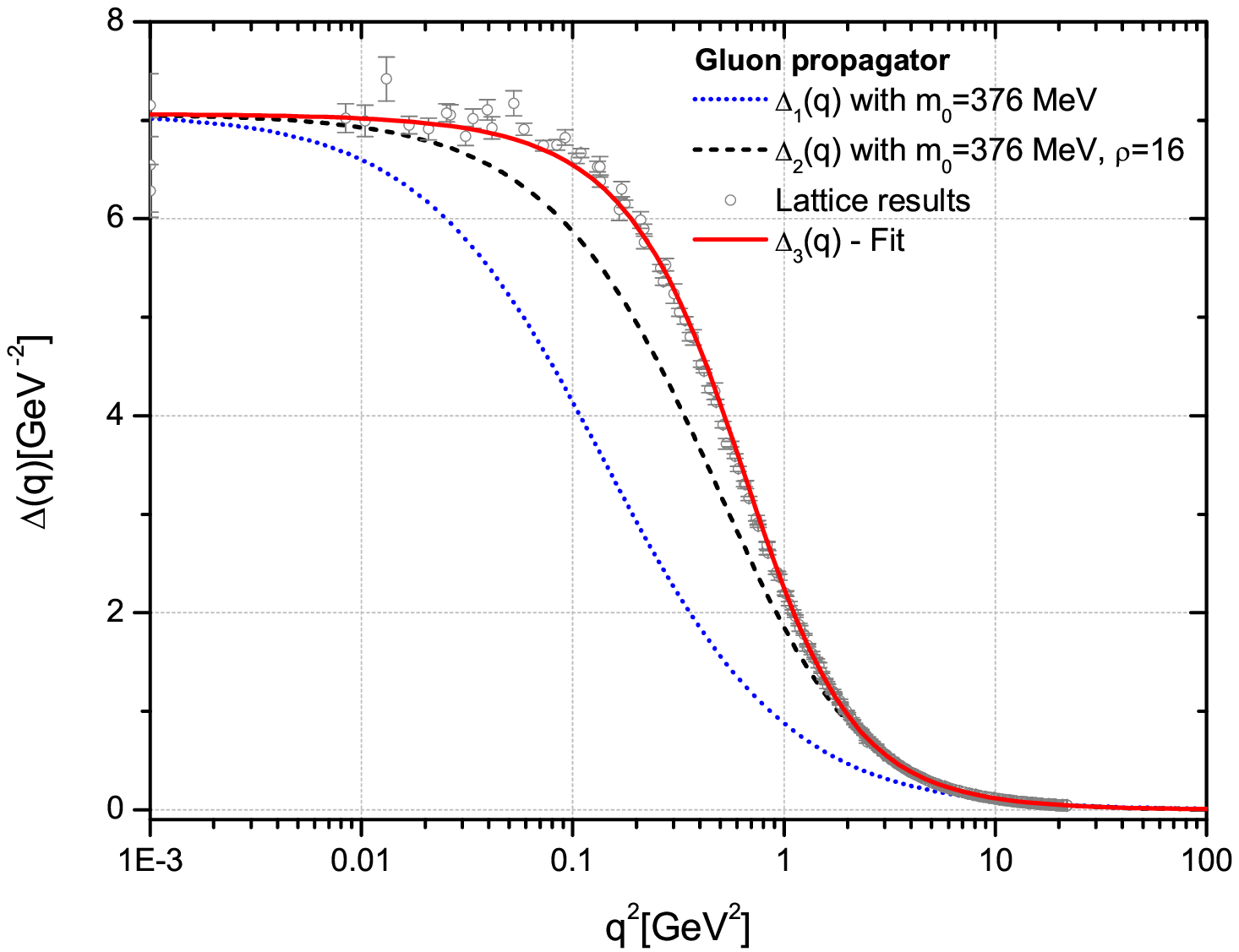}
\end{minipage}
\begin{minipage}[b]{0.50\linewidth}
\hspace{-0.5cm}
\noindent
\includegraphics[scale=0.55]{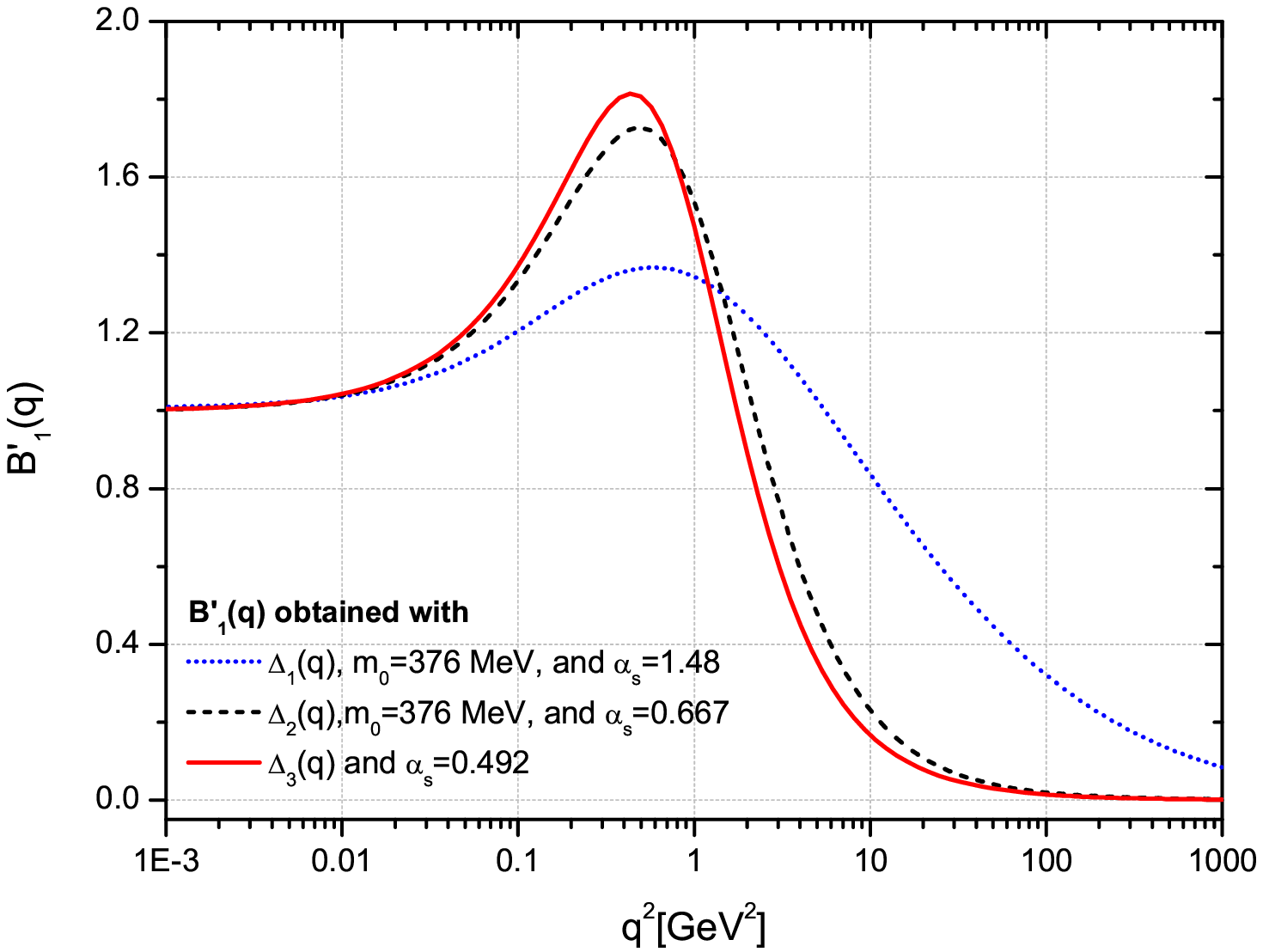}
\end{minipage}
\caption{The three models for the gluon propagator (left) and 
the corresponding solutions of the BS equation for $B'(x)$ (right)}
\label{props}
\end{figure}

{\bf (i)} Let us start with the simplest such propagator, namely a tree-level massive propagator of the form 
\be
\Delta_1^{-1}(q^2) = q^2 + m^2_0 \,,
\label{smassive}
\ee 
where $m^2_0$ is a hard mass, that will be treated as a free parameter. 
On the left panel of Fig.~\ref{props}, the (blue) dotted curve represents 
$\Delta_1(q^2)$ for $m_0=376 \,\mbox{MeV}$.

{\bf (ii)} The second model is an improved version of the first, where 
we introduce the renormalization-group logarithm next to the momentum $q^2$, more specifically
\be
\Delta_2^{-1}(q^2)=  q^2\left[1+ \frac{13C_{\rm A}g^2}{96\pi^2} 
\ln\left(\frac{q^2 +\rho\,m_0^2}{\mu^2}\right)\right] +  m^2_0 \,,
\label{gluon2}
\ee  
where $\rho$ is an adjustable parameter varying in the range of $\rho \in [2,10]$. Notice that the hard mass $m_0^2$
appearing in the argument of the perturbative logarithm  
acts as an infrared cutoff; so, instead of the logarithm diverging at the Landau pole, it
saturates at a finite value. The (black) dashed line
represents the Eq.~(\ref{gluon2}) when $\rho=16$, $m_0=376 \,\mbox{MeV}$, and  $\mu=4.3$ GeV.

{\bf (iii)} The third model is simply a 
physically motivated fit for the gluon propagator determined
by the large-volume lattice simulations of Ref.~\cite{Bogolubsky:2007ud}, and 
shown on the left panel of Fig.~\ref{props}. 
The lattice data presented there correspond to a $SU(3)$  quenched 
lattice simulation, where $\Delta(q)$ is renormalized at $\mu=4.3$ GeV.  
This gluon propagator 
can be  accurately fitted by the expression 
\be
\Delta_3^{-1}(q^2)= m_g^2(q^2) + q^2\left[1+ \frac{13C_{\rm A}g_1^2}{96\pi^2} 
\ln\left(\frac{q^2 +\rho_1\,m_g^2(q^2)}{\mu^2}\right)\right]\,,
\label{gluon3}
\ee  
where  $m_g^2(q^2)$ is a running mass given by
\be
m^2_g(q^2) = \frac{m^4}{q^2 + \rho_2 m^2} \,,
\label{dmass}
\ee
and the values of the fitting parameters are  
\mbox{$m= 520$\,\mbox{MeV}}, \mbox{$g_1^2=5.68$}, \mbox{$\rho_1=8.55$} and, \mbox{$\rho_2=1.91$}.
On the left panel of Fig.~\ref{props}, the (red) continuous line represents
the fit for the lattice gluon propagator given by Eq.~(\ref{gluon3}). Notice that, in all three cases, we 
have fixed the value of $\Delta^{-1}(0)=m_0^2\approx 0.14$.

Our main findings may be summarized as follows.

{\it (a)} In Fig.~\ref{props}, right panel, we show the solutions 
of Eq.~(\ref{weakBS}) obtained using as input 
the three propagators shown on the left panel.
For the simple massive propagator
of Eq.~(\ref{smassive}), a solution for $B'_1(q)$ is found for \mbox{$\alpha_s=1.48$};
in the case of $\Delta_2(q)$ given by  Eq.~(\ref{gluon2}), a solution 
is obtained when \mbox{$\alpha_s=0.667$}, while for the lattice
propagator $\Delta_3(q)$ of  Eq.~(\ref{gluon3}) a non-trivial 
solution is found when \mbox{$\alpha_s=0.492$}.

{\it (b)} Note that, due to the fact that Eq.~(\ref{weakBS}) is homogeneous and (effectively) linear,
if $B'_1(q)$ is a solution then the function $cB'_1(q)$ is also a solution, 
for any real constant $c$. Therefore,
the solutions shown on the right panel of Fig.~\ref{props} corresponds to  
a representative case of a family of  possible solutions, where the constant $c$ was chosen 
such that $B'_1(0)=1$.

{\it (c)} Another interesting feature of the solutions of Eq.~(\ref{weakBS}) is  
the dependence of the observed peak on the support of the gluon propagator 
in the intermediate region of momenta. Specifically, 
an increase of the support of the gluon propagator in the approximate range (0.3-1) GeV 
results in a more pronounced peak in  $B'_1(q)$.

{\it (d)} In addition, observe that due to the presence of the 
perturbative logarithm 
in the expression for $\Delta_2(q)$ and $\Delta_3(q)$, the corresponding solutions $B'_1(q)$  
fall off in the ultraviolet region much faster than those obtained using the 
simple $\Delta_1(q)$ of Eq.~(\ref{smassive}). 

\section{Conclusions}

In this presentation we have 
reported recent progress~\cite{arXiv:1110.2633} on the study of the Schwinger
mechanism in QCD, which is the only self-consistent way to 
endow gluons with a dynamical mass. 
This mechanism relies on the existence of massless bound-state excitations, 
whose dynamical formation is controlled by a homogeneous BSE.
As we have seen, under certain simplifying assumptions, this 
equation admits non-trivial solutions, thus furnishing additional 
support in favor of the specific mass generation mechanism described 
in a series of earlier works~\cite{Aguilar:2006gr,Aguilar:2008xm,arXiv:1107.3968}. 

In the future it would 
be particularly important  to consider the effects of  
bound-state poles in the SD kernels of not only the three-gluon vertex, 
as we did here, but of all other 
fundamental  vertices  of  the  theory. 
Such  an  investigation  would eventually give rise 
to a coupled system of various homogeneous integral equations.
Especially interesting  in this  context is the  information
that  one  might be  able  to obtain  on the  
corresponding wave-function of the ghost-ghost channel.
Specifically, according to the recent lattice 
findings~\cite{Cucchieri:2007md,Cucchieri:2009zt,Bogolubsky:2007ud,Bowman:2007du,Bogolubsky:2009dc}, 
in the deep infrared 
the ghost dressing  function $F$ 
is finite, but the full ghost propagator diverges, a fact that strongly 
suggests that there  is  no
dynamical  mass associated with  the ghost  field (note that  
the finiteness of $F$ can be easily accounted for by the  
presence of a gluon mass, saturating the perturbative logarithm of $F$~\cite{Aguilar:2008xm}).
One  would expect,
therefore, that  the solution of the corresponding  system should give
rise to a non-vanishing $B_1'$, as before, but to a vanishing 
ghost-ghost wave function.

\newpage 

{\it Acknowledgments:} 

I would like to thank the ECT* for making the QCD-TNT II workshop possible. 
This research was supported by the European FEDER and  Spanish MICINN under 
grant FPA2008-02878.

\end{document}